\documentclass[aps,prl,reprint,superscriptaddress,showpacs]{revtex4-1}
\usepackage{graphicx}  

\begin{document}
\title{Dynamical I-V Characteristics of SNS Junctions} 
\author{Kevin Spahr}
\affiliation{Institut Quantique, D\'{e}partement de Physique, Universit\'{e} de Sherbrooke, Sherbrooke, Qu\'{e}bec, Canada, J1K 2R1}
\author{Jonathan Graveline}
\affiliation{Institut Quantique, D\'{e}partement de Physique, Universit\'{e} de Sherbrooke, Sherbrooke, Qu\'{e}bec, Canada, J1K 2R1}
\author{Christian Lupien}
\affiliation{Institut Quantique, D\'{e}partement de Physique, Universit\'{e} de Sherbrooke, Sherbrooke, Qu\'{e}bec, Canada, J1K 2R1}
\author{Marco Aprili}
\affiliation{Laboratoire de Physique des Solides (CNRS UMR 8502), Universit\'{e} Paris-Saclay, 91405 Orsay, France}
\author{Bertrand Reulet}
\affiliation{Institut Quantique, D\'{e}partement de Physique, Universit\'{e} de Sherbrooke, Sherbrooke, Qu\'{e}bec, Canada, J1K 2R1}

\date{\today}

\begin{abstract}
We have probed the switching dynamics of the Josephson critical current of a superconducting weak link by measuring its voltage/current characteristics while applying an ac current bias in the range 1-200 MHz. The weak link between two Nb reservoirs is formed by an mesoscopic Al wire above its critical temperature. We observe a dynamical phase transition as a function of the frequency and amplitude of the ac current. While at low frequency the transition driven by increasing the current bias is well described by the standard Kramers theory, at high frequency the switching histograms become hysteretic and much narrower than expected by thermal fluctuations. The crossover frequency between the two regimes is set by the electron-phonon interaction rate in the normal metal. 

\end{abstract}

\maketitle

A thermodynamic phase transition describes the change of state of a macroscopic ensemble of particles under an adiabatic variation of an external parameter. For example certain electrical conductors go from a resistive state (i.e., a normal conductor N) to a dissipationless state (a superconductor S) below a critical temperature. Similarly, a Josephson junction (JJ) made of two S contacts weakly connected, switches from a zero resistance state to a resistive one when driven by a dc current larger than a critical current $I_c$. A dynamical phase transition occurs when the system is driven by a parameter that changes quickly enough so that the system does not have time to equilibrate. Here we investigate such a dynamical phase transition in a Superconductor / Normal metal / Superconductor junction (SNS, i.e. a JJ where the weak link is made of a normal metal) driven by a time-dependent current whose amplitude and frequency are  larger than $I_c$ and  the energy relaxation time in N, respectively.

The transition from non-dissipative (zero resistance) to dissipative (finite resistance) regimes in a JJ corresponds to a change in dynamics of the phase difference $\varphi$ between the superconducting contacts \cite{Tinkham}. $\varphi$ either stays in a local minimum of the washboard potential describing the junction energy (non-dissipative state, for a current $I<I_c$) or runs down (dissipative state, $I>I_c$). Thermal fluctuations promote escape from the local minima \cite{Kurkijarvi}. Therefore at finite temperature switching from the phase locked to the running state is reached before the bias current equals the critical current \cite{Buttiker}. By driving the junction with a slow, periodic current ramp one can construct switching histograms that measure the thermal escape rate  \cite{Fulton} usually given by the Kramers theory \cite{Kramers}. If the phase follows adiabatically the excitation, the width of the histogram is simply related to the phase temperature. Alternatively, if the current bias changes faster than the phase relaxation rate $\gamma_\varphi$, a non-adiabatic regime appears  \cite{Dykmann,Stambaugh} resulting in a dynamical phase transition \cite{Petkovic}.

 In underdamped  junctions, incomplete energy relaxation after switching leads to a dynamical phase bifurcation \cite{Dykmann2}. This bifurcation manifests itself as premature switching \cite{Dykmann2, Nayfeh}, resulting in a bimodal switching distribution when the current ramp frequency is of the order or higher than $\gamma_\varphi$ \cite{Petkovic}. In junctions in which energy quantization is larger than the bath temperature (quantum limit), refilling of the top energy levels from the lowest by thermal diffusion is prevented for a sweep rate higher than $\gamma_\varphi$ \cite{Barone}. As empty levels do not contribute to the Kramers escape, this out-of-equilibrium energy distribution is revealed in the switching histograms \cite{Russo}.

In SNS junctions the phase dynamics is overdamped  \cite{Song} and hence the dynamics is not set by circuit parameters and electric elements as in the case of  Superconductor / Insultaor / Superconductor junctions (the tunneling time  can be neglected being $<<1/ \gamma_\varphi$), but instead by intrinsic microscopic time scales  \cite{Virtanen} : the diffusion rate of electrons along the N part, $\gamma_D$, and the energy relaxation rate in N, $\gamma_E$, originating from electron-phonon interactions. These two characteristic time scales have been shown to set the critical and the re-trapping currents respectively, measured in the dc $I(V)$ characteristics  in absence  \cite{Dubos,Angers} and  in  presence \cite{Chiodi} of an ac excitation.

Here we confirm these results and we show that it is the energy relaxation rate in N, $\gamma_E$ which governs the dynamics of the phase transition. To do so we have measured the  $V(I)$ characteristics of the junction at finite frequency, $f \approx \gamma_E$ and recorded the S-to-N switching  and the N-to-S switching back in the time domain when the SNS junction is biased by a radiofrequency current $I(t)=I_0\cos2\pi ft$. 

\begin{figure}[htb]
		\includegraphics[width=0.9\columnwidth]{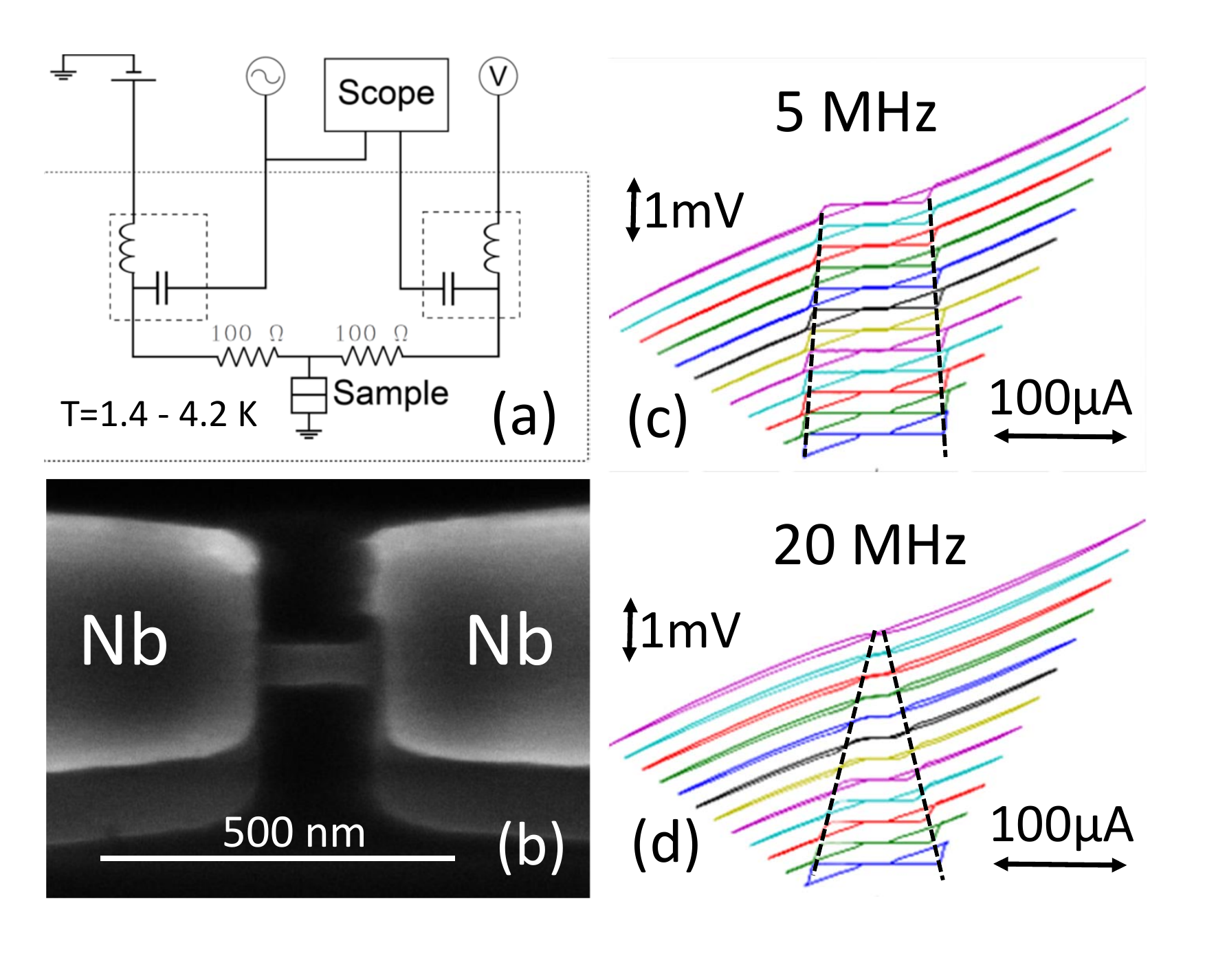}
\caption{(a) Experimental setup. The $100\Omega$ resistor on the left of the sample is used to insure proper ac current bias while that on the right increases the input impedance of the amplifier to make it a good ac voltmeter. (b) SEM image of the sample. (c) $V(I)$ for various amplitudes of the excitation current at frequency $f=5$ MHz. (d) idem at $f=20$ MHz. $V(I)$ curves in (c) and (d) have been shifted vertically for clarity. The dashed black lines indicate the position of the S-to-N switching.}
    \label{fig:setup}
\end{figure}

\emph{Experimental setup.} 
We have prepared SNS short junctions by angle evaporation. A 495PMMA/950PMMA mask has been obtained by electron lithography. 10nm of Al (here N) and then 50nm of Nb (S) have been subsequently evaporated in UHV (base pressure  $10^{-9}$mBar). Shadow evaporation of Nb, the sample being tilted at $36^\circ$, prevents Nb to be deposited on the Al bridge. A SEM picture of a sample after lift-off in NMP (N-Methyl-2-Pyrrolidon) is presented in Fig. \ref{fig:setup}(b). The critical temperature of Nb is about $7.5$K,  $R_n \approx 7\Omega$ is the normal resistance of the Al bridge at $4.2$K, $L \approx 300$nm and $D \approx 70$cm$^2$/s. The critical current $I_c$ at $4.2$K is a few microamps and about $50\mu\mathrm{A}$ at $1.6$K. Consistently with previous experiments, we have observed an hysteresis in the $V(I)$ characteristics of the junctions below 2K \cite {Chiodi}.

 The experimental setup used to measure the $V(I)$ characteristics at finite frequency is illustrated on Fig. \ref{fig:setup}(a). The sample has been placed in a pumped $^4$He cryostat  and the temperature adjusted above the Al critical temperature, between 1.4 K and 4.2 K. A $100 \Omega $ resistor has been connected in series with a $50\Omega$ RF voltage source to ensure current bias of the sample ($100+50\gg R_n$). Similarly, a $100\Omega$ resistor  has been connected in series with the $50\Omega$ detector to ensure a proper voltage detection ($100+50\gg R_n$). Both resistors have been soldered close to the sample (at a few mm), i.e. at a distance that is always much smaller than the wavelength of the excitation. This provides a flat response function of the setup up to $\sim 1$GHz. The voltage across the sample is amplified by a cryogenic amplifier, then detected at room temperature. Examples of $V(I)$ characteristics measured with an oscilloscope are shown in Figs. \ref{fig:setup}(c) and \ref{fig:setup}(d). At radiofrequencies, the propagation delays cannot be neglected, so there is a large lag between the voltage at the RF source and the current in the sample as well as between the voltage across the sample and that observed on the oscilloscope. The overall delay has been calibrated out by driving the SNS junction with a large dc current. In the normal state the junction behaves as a resistor with a purely dissipative impedance. Bias-tees on each side of the junction have been added to measure $V(I)$ at zero frequency (dc). Their crossover frequency $\sim0.1$MHz is low enough not to perturb the measurements at high frequency.  

\emph{Results}.
We present in Figs. \ref{fig:setup}(c),(d) the $V(I)$ characteristics of the junction measured using an ac current bias $I(t)=I_0\cos(2\pi ft)$ for two frequencies $f=5$MHz (c) and $f=20$MHz (d) and different amplitudes $I_0$. The curves have been shifted vertically for the sake of clarity. We observe that $V(I)$ curves  overlap at low frequency, meaning that the $V(I)$ characteristics is independent of $I_0$ (the dashed lines that indicate the position of the S-to-N switching are almost vertical). This situation is almost identical to the zero frequency measurement and corresponds to an adiabatic situation for the junction. In contrast, the $V(I)$ curves measured at higher frequency clearly depend on the maximum applied current $I_0$ (the dashed lines tend to join each other at high $I_0$). While for $I_0$ slightly larger than the critical current, the observed $V(I)$ is very close to the dc one, deviations from this behavior become stronger and stronger when $I_0$ is increased, leading to the modification of the $V(I)$ characteristics and a suppression of the supercurrent and different voltages for increasing and decreasing bias. Note that there is an hysteresis in the $V(I)$ characteristics even at zero frequency, and is related to quasiparticle heating \cite {Courtois,DeCecco} : above $I_c$ the electrons in the junction are heated by the dissipated Joule power up to a temperature $T_e>T_{ph}$ with $T_{ph}$ the bath temperature of phonons. Thus the critical current is lowered and one has to decrease the current below $I_c(T_{e})$ to drive the junction back to the dissipationless state.

\begin{figure}[htb]
		\includegraphics[width=0.8\columnwidth]{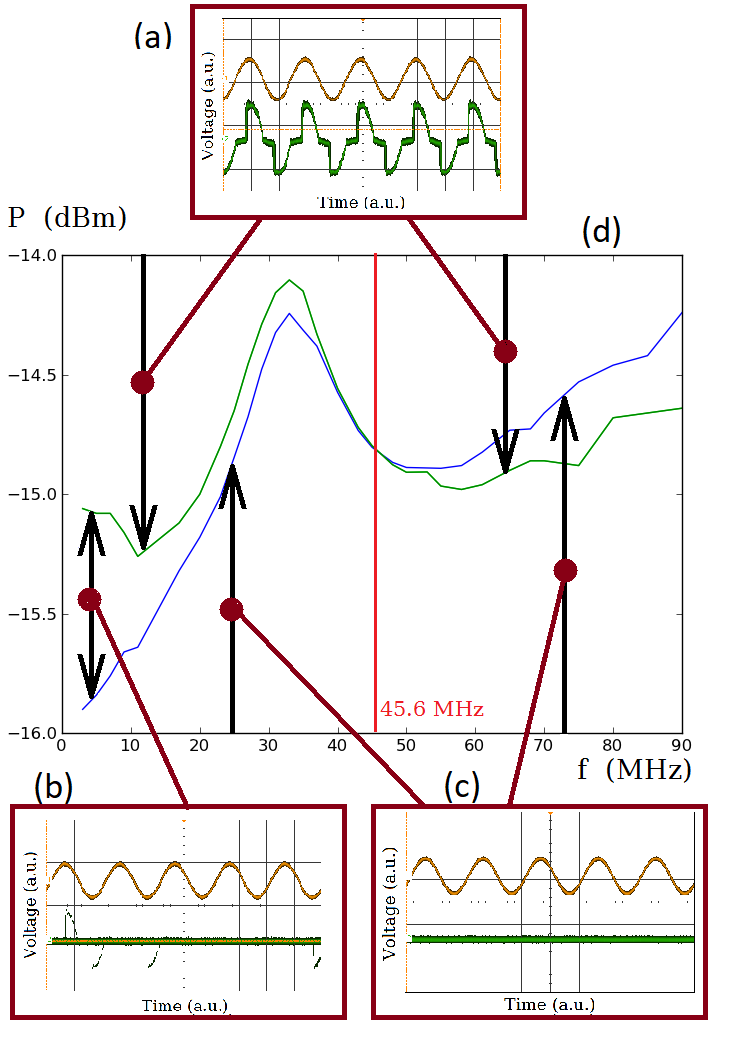}
    \caption{(a)-(c): screenshots of the oscilloscope for different excitation power and frequency. The upper curve (yellow) is $I(t)$, the lower one (green) is $V(t)$. (d) The lines represent the change in the dynamical behavior of the junction as a function of frequency: the blue line represent the upper limit of the Dyn-S state ($I_0<I_S$), the green one the lower limit of the Dyn-N state ($I_0>I_N$). The red vertical line corresponds to the disappearance of the fluctuating regime at $f=f^*$.}
    \label{fig:diag_phase}
\end{figure}

The $V(I)$ characteristics presented in Figs. \ref{fig:setup}(c),(d) result from averaging over thousands of cycles. One may wonder whether all individual cycles are identical up to the experimental noise, or they actually correspond to different transients as expected when thermal fluctuations lead to a  Brownian motion of the phase. 
To investigate the dynamics of the S-to-N transition we present in Fig. \ref{fig:diag_phase}(a)-(c) screenshots of the oscilloscope, i.e. time traces of $I(t)$ (upper curves) and $V(t)$ (lower curves) on time scales of a few periods of the excitation and $I_0\approx I_c$ . Depending on excitation current and frequency, the junction behaves differently. Sequential switching events are recorded over long periods, here over $200\mu$s (the effect of the measurement time on our results will be discussed later).
At low enough excitation power, Fig. \ref{fig:diag_phase}(c), we observe that $V(t)=0$. We will call this regime the dynamical S state (Dyn-S). In this regime $I_0$ is too weak and transport through the junction  is always dissipationless. We define $I_S(f)$ as the highest amplitude of the bias current that the junction can sustain while in  Dyn-S  within the recording period. The blue curve in Fig. \ref{fig:diag_phase} is that of $I_S(f)$ \cite{peak}.
 At higher microwave power, Fig. \ref{fig:diag_phase}(a), the junction switches S-to-N and back N-to-S at each cycle, regardless of frequency. 
We will call this regime the dynamical N state (Dyn-N) and define $I_N(f)$ the smallest value of $I_0$ such that the junction is in  Dyn-N . The green curve Fig. \ref{fig:diag_phase}(d) is that of $I_N(f)$.
Finally for intermediate power we observe two distinct behaviors as a function of the frequency of the excitation. At low frequency in the region of parameters that are between the blue and green lines, i.e. $I_S(f)<I_0<I_N(f)$, switching is uncertain and the probability that the junction switches to the normal state at each cycle goes from zero to one by increasing power, see Fig. \ref{fig:diag_phase}(b). The stochastic origin of switching is related to thermal fluctuations as discussed above. However,  as it can be seen in Fig. \ref{fig:diag_phase}(d), this region shrinks when the frequency is increased, up to a critical value $f^*=45.6$MHz where $I_S=I_N$. At this critical value stochasticity disappears:  the junction is always in the superconducting state for $I_0<I_S$ and it switches always from S to N at each cycle once the excitation current is increased. For $f>f^*$ the fluctuating regime is replaced by an hysteretic transition between these two dynamical states: starting from $I_0<I_S$ and increasing the power, the transition from Dyn-S to Dyn-N  is abrupt (blue line of Fig. \ref{fig:diag_phase}(d)). When the power is decreased the transition back from Dyn-N to Dyn-S  is also abrupt but it occurs at a lower current (green line). Note that $f^*=45.6$MHz corresponds exactly to the value measured in similar junctions for the transition from two-steps to one-step $I(V)$ characteristics  and it is comparable with the cut-off frequency due to the  energy relaxation time  in Al bridges at this temperature \cite{Chiodi}, \cite{Prober}.

\begin{figure}[htb]
		\includegraphics[width=1.5\columnwidth]{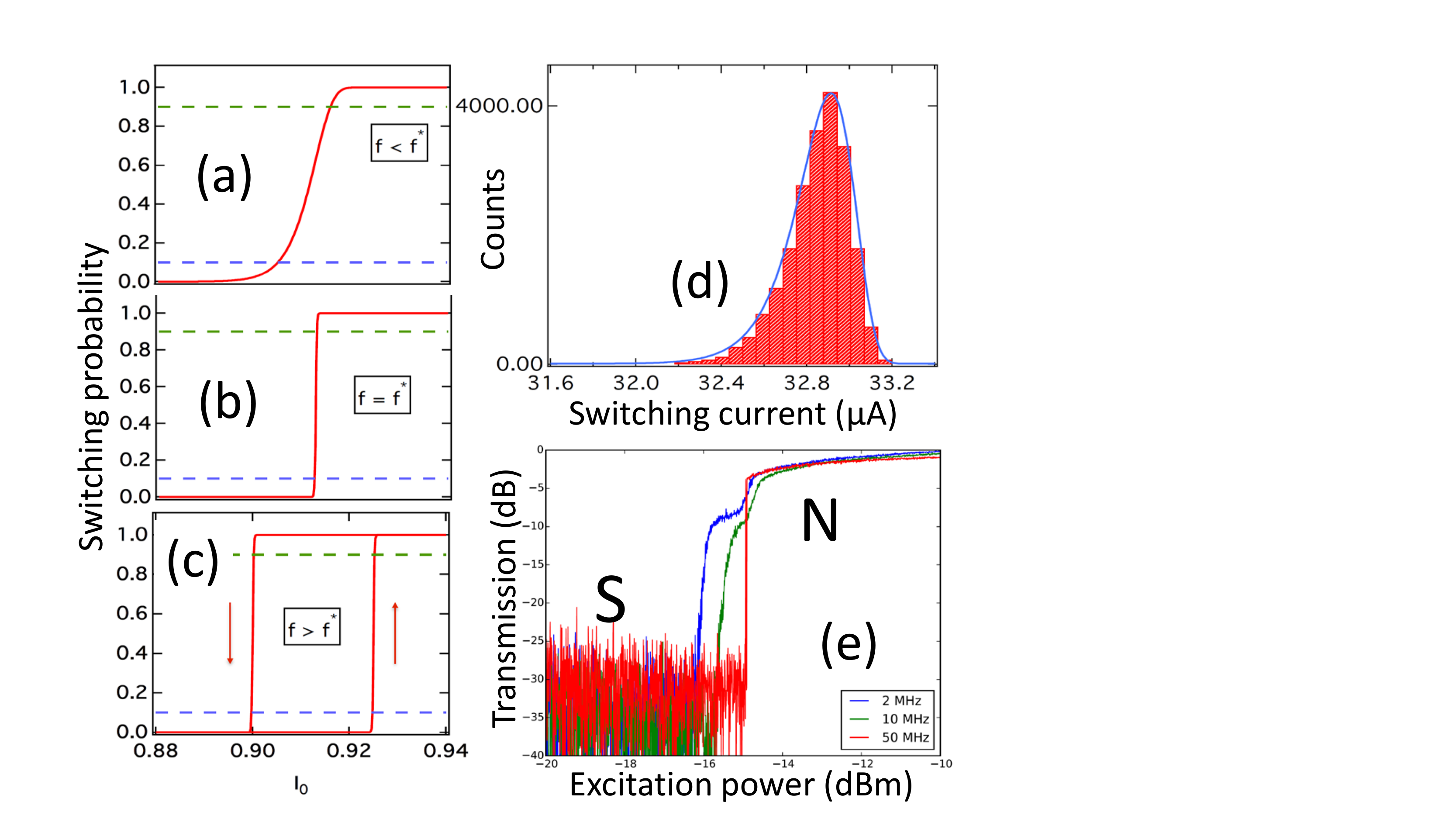}
    \caption{Schematics of the switching probability $P$ as a function of the excitation $I_0$ for low frequency $f<f^*$ (a), at the cross-over $f=f^*$ (b) and in the high frequency, hysteretic regime $f>f^*$ (c). The blue line represents $P(I_0$)=0.1, the green one $P(I_0)=0.9$. (d): switching histograms measured at very low frequency $f\ll f^*$. (e): RF transmission of the setup as a function of the excitation power for various frequencies.}
    \label{fig:cartoon}
\end{figure}
 
To illustrate these observations, we have sketched in Fig. \ref{fig:cartoon}(a)-(c) a cartoon that represents the evolution of the switching probability $P$  from S to N during a half cycle, as a function of $I_0$ when the frequency of the bias current is increased. In the adiabatic regime $f\ll f^*$, i.e. when the bias changes slower than both the phase relaxation time and the  energy relaxation time, the escape rate is given by the Kramers decay \cite{Kramers} due to thermal fluctuations. The probability to switch from S to N depends mainly on how close the bias current is to the critical current and on the phase temperature of the junction. This gives rise to a probability integrated over a half period of the sine wave excitation that is sketched in Fig. \ref{fig:cartoon}(a). The blue and green dashed lines represent $P(I_0)=0.1$ and $P(I_0)=0.9$ which correspond to the blue ($I_S$) and green ($I_N$) lines of the dynamical phase diagram of Fig. \ref{fig:diag_phase}(d). We have measured switching histograms at low frequency $f_0 = 33$Hz $\ll f^*$. An example of such histogram is shown in Fig. \ref{fig:cartoon}(d) together with the theoretical prediction \cite{Garg95} (solid line), which coincide, thus indicating the absence of spurious noise sources in our experiment. At low frequency $P(I_0)$ is simply the integral from zero to $I_0$ of the normalized switching histograms. While the regimes  $P=0$ (Dyn-S state, $I_0<I_S$) and $P=1$ (Dyn-N state, $I_0>I_N$)  persist at frequency $f>f^*$, the shape of $P(I_0)$ depends strongly on the bias frequency. When the frequency is increased, $I_N$ and $I_S$ converge so that  $P(I_0)$ becomes steeper  until $f=f^*$ where  $I_N=I_S$ and $P(I_0)$ is a step function as sketched in Fig. \ref{fig:cartoon}(b). In other words, the switching probability $P$ can only be either 0 or 1 and the junction switches always or never at each cycle. To demonstrate experimentally the effect of an increased excitation frequency on $P$ we have measured the RF transmission through the sample as a function of the excitation power for various frequencies, see Fig. \ref{fig:cartoon}(e). When the sample is superconducting, it behaves as a short circuit and almost no power is transmitted; when it is normal, the transmission in finite. Curves in Fig. \ref{fig:cartoon}(e) are averaged, so they reflect the switching probability $P$. One clearly observe that the curves become steeper when the frequency increases: at frequencies up to 30 MHz the curves are continuous, meaning that the S to N switching occurs not always at the same value; for frequencies above 50 MHz there is an abrupt step, which means that the S-to-N transition occurs always for the same $I_0$. Finally for higher frequency $f>f^*$, $I_S$ keeps increasing while $I_N$ keeps decreasing, see Fig. \ref{fig:diag_phase}(d). This corresponds to the appearance of an hysteresis in the dynamical behavior of the junction: for $I_N<I_0<I_S$ the junction can be either in Dyn-S or Dyn-N  depending on its history \cite{peak}. This regime is sketched in Fig. \ref{fig:cartoon}(c).

\begin{figure}[htb]
		\includegraphics[width=0.5\textwidth]{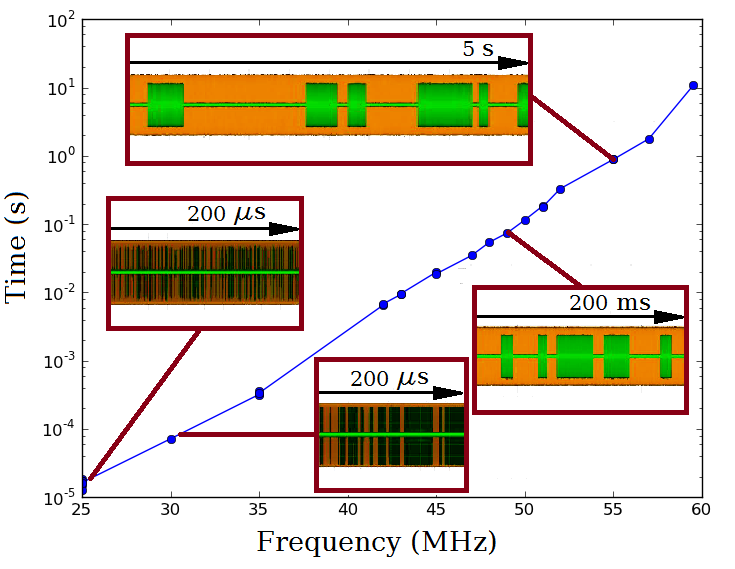}
    \caption{Correlation time $\tau_c$ (log scale) as a function of the excitation frequency $f$ (linear scale) at $T=1.40$K. Insets: oscilloscope time traces with time scales.}
    \label{fig:bunch}
\end{figure}

\emph{Correlations.}
We now consider the following question: are switching events independent of each other ? To answer that question we have recorded $V(t)$ traces over very long times, from $200\mu$s to $5$s (i.e., $>10^8$ cycles \cite{spectre}). Examples of such traces are shown inside the plot of Fig. \ref{fig:bunch} (green / black curves on orange / brown background) for different frequencies between $25$ and $60$MHz. A small signal means that the junction remained superconducting and acted as a short circuit, a larger signal  instead means that the junction has switched during the period. We observe that while at $f=25$MHz the switching events are independent, they become correlated by increasing the bias frequency : the junction can remain in Dyn-N or Dyn-S  for more than a second at $f=55$MHz. This implies correlation among hundreds of millions of consecutive cycles. To be more quantitative, we have made histograms of the durations of the "bunching" of cycles and determined their average $\tau_c$. In order to determine $\tau_c$ as a function of frequency while approaching $f^*$ we chose to adjust $I_0$ for each frequency such that the junction switches half of the time in average ($P=1/2$) at each frequency. This corresponds to scan the dynamical phase diagram of Fig. \ref{fig:diag_phase} while staying in the middle between the blue and green lines. We report in Fig. \ref{fig:bunch} the result of $\tau_c(f)$. We clearly observe that the correlation time diverges exponentially (over 6 decades) when $f$ increases, i.e.: $\tau_c\propto\exp(f\tau_*)$ with $\tau_*^{-1}\sim2.5$MHz. This phenomenon resembles the slow-down observed near a critical point except that here we do not observe a power law in $|f-f^*|$ but rather a more rapid increase of $\tau_c$ with frequency.

This results sheds light on the effect of the measurement time on the phase diagram of Fig. \ref{fig:diag_phase}, which has been obtained using a measurement time of $200\mu$s.
Using longer acquisition time in the fluctuating regime would decrease $I_N$ and increase $I_S$: one could see e.g. no switching during $200\mu$s but see one had one waited longer. Taking longer measurement time would also increase $f^*$.  

\emph{Discussion.}
At low frequency the distribution of the switching current is known to be related to phase fluctuations of the junction which, except at very low temperature, are of thermal origin. Our low frequency data are indeed well explained by Kramer's theory. The dynamical properties of SNS junctions involve the phase dynamics but also the dynamics of energy exchange between the electrons in the normal region and the phonons in the substrate. A theoretical description of the dynamical switching of an SNS junction should thus couple phase and thermal dynamics \cite{DeCecco}. This can be done  phenomenologically by introducing a time-dependent electron temperature $T_e(t)$ and a temperature dependent critical current $I_c(T_e)$ in the Resistive-Shunted-Junction (RSJ) model \cite{Tinkham}. In our experiment the frequency is always much smaller than $1/\tau_D$ and hence of the mini-gap in N, therefore enhancement of $I_c$ by microwave pumping \cite{Virtanen, Chiodi} can be easily ruled out. As consequence, the dissipation in the junction can be mainly accounted for an electron temperature (the electron-electron scattering time being much shorter than the electron-phonon scattering time, it is legitimate to consider that the temperature is well defined on the time scale discussed here). 

Fluctuations which might be at the origin of "bunching" as reported in Fig. \ref{fig:bunch} can be added in two ways. First, current fluctuations are naturally introduced in the RSJ model by  adding a Langevin term due to the quasiparticle thermal noise. Second, the normal part of the junction is both thermally isolated (heat can leave only by phonon emission) and of very small volume. Thus its energy and temperature fluctuate in time, on a time scale given by the electron-phonon interaction time, of the order of 10ns. An estimate of the variance of its thermal fluctuations is given by $\langle\Delta T^2\rangle\sim k_BT^2/C$ with $C$ the heat capacity, proportional to the volume of the N part (this is valid at equilibrium only, temperature fluctuations are greater in the presence of Joule heating). We find $\sqrt{\langle\Delta T^2\rangle}\sim22$mK. If the critical current follows instantaneously the temperature variations, this corresponds to typical critical current variations of $0.3\mu$A, very similar to what we observe. Clearly temperature fluctuations might play an important role here, as they do for example in limiting the noise equivalent power of nanobolometers \cite{nanobolo}. In this picture the temperature fluctuations are due to phonon shot noise, and the divergence of the correlation time would be associated to very rare events of emission or absorption of very energetic phonons (or many succesive phonons) which induce a large enough temperature variation that provokes the switching of the junction between Dyn-N and Dyn-S states. Such events are indeed exponentially rare. 

\emph{Conclusion.}
We have reported measurements that probe the dynamics of the Josephson couping in mesoscopic SNS junctions. We have observed a dynamical phase transition as a function of frequency and excitation power, with a characteristic frequency similar to the electron-phonon rate in accordance with previous experiments \cite{Chiodi}. These measurements raise the question whether  the dynamical  transition from the superconducting to the normal state can be accurately described by the dynamics of an effective average electron temperature in the normal metal or if temperature fluctuations should be also taken into account. More experimental and theoretical works are needed to understand the interplay between phase and temperature dynamics,  and noise in SNS structures.

We are very grateful to F. Bintou-Sane, J. Gabelli and D. Prober for fruitful discussions. This work has been supported by the ANR Blanc grant (DYCOSMA) from the French Agence Nationale de Recherche, the Canada Excellence Research Chairs, Government of Canada, Natural Sciences and Engineering Research Council of Canada, Qu\'ebec MEIE, Qu\'ebec FRQNT via INTRIQ, Universit\'e de Sherbrooke via EPIQ, and Canada Foundation for Innovation.

%


\begin{thebibliography}{99}
\bibitem{peak} We observe in Fig. \ref{fig:diag_phase} a maximum in both $I_N$ and $I_S$ at $f\sim33$MHz. We have observed similar maxima in other samples at different frequencies, with no obvious link between the parameters of the sample and the frequency of the maximum and no temperature dependence. Even though we tried to have a setup with a flat frequency response, we cannot rule out that the observed maximum is due to a resonance of $\sim0.8$dB in the experimental setup. The ratio $I_N/I_S$, which is probably much less sensitive to experimental imperfections, does not show any structure in its frequency dependence.
\bibitem{spectre} Recording traces over that many cycles would require a memory well beyond that of our oscilloscope. Thus we measured the power transmitted through the junction at the excitation frequency averaged over one period. A small power means a superconducting junction while a large power means that the junction has switched to the normal state. Thus one period corresponds to only one number to record. This has been performed using a spectrum analyzer in zero span mode.
\end{thebibliography}
\end{document}